\documentclass[apj,numberedappendix]{emulateapj}
\usepackage{epsfig}
\usepackage{amsmath}
\usepackage{enumerate}
\usepackage{natbib}
\usepackage{hyperref}

\def\sT{\sigma_{\rm T}}
\def\beq{\begin{equation}}
\def\eeq{\end{equation}}
\def\Ep{E_{\rm pk}}
\def\RP{R_{\rm P}}
\def\ThP{\Theta_{\rm P}}
\def\TP{T_{\rm P}}
\def\TW{T_{\rm W}}
\def\tauP{\tau_{\rm P}}
\def\RW{R_{\rm W}}
\def\Rph{R_\star}
\def\Rsat{R_{\rm s}}
\def\Rdiss{R_{\rm diss}}
\def\Th{\Theta}
\def\dM{\dot{M}}
\def\dN{\dot{N}_\gamma}
\def\gampeak{\gamma_{\rm peak}}
\def\Eav{\bar{E}}
\def\Sect{Section}
\def\Sects{Sections}
\def\Eq{Equation}
\def\Eqs{Equations}
\def\texp{t_{\rm exp}}
\def\Urad{U_\gamma}
\def\Trad{T_\gamma}

\def\Lrad{L_\gamma}
\def\ep{\epsilon}
\def\nP{n_{\rm P}}
\def\dnB{\dot{n}_{\rm B}}
\def\dnDC{\dot{n}_{\rm DC}}
\def\dn{\dot{n}}
\def\lbar{\lambda\llap {--}}
\def\xmin{x_{\min}}
\def\tIC{t_{\rm IC}}
\def\tabs{t_{\rm abs}}
\def\aDC{a_{\rm DC}}

\def\tC{t_{\rm C}}
\def\thb{\theta_b}
\def\Rcoll{R_{\rm coll}}
\def\etaB{\eta_B}
\def\etath{\eta_{\rm th}}
\def\etaP{\eta_{\rm P}}
\def\etaW{\eta_{\rm W}}
\def\tauW{\tau_{\rm W}}
\def\GamP{\Gamma_{\rm P}}
\def\GamW{\Gamma_{\rm W}}
\def\ThW{\Theta_{\rm W}}
\def\EavW{\Eav_{\rm W}}
\def\EavP{\Eav_{\rm P}}
\def\epP{\ep_{\rm P}}
\def\epW{\ep_{\rm W}}
\def\epnth{\ep_{\rm nth}}
\def\Eabs{E_{\rm abs}}

\def\dMacc{\dot{M}_{\rm acc}}
\def\dm{\dot{m}}
\def\xms{x_{\rm ms}}
\def\Teff{T_{\rm eff}}
\def\fKN{f_{\rm KN}}
\def\Ein{E_0}
\def\Emin{E_{\min}}
\def\Lacc{L_{\rm acc}}

\newbox\grsign \setbox\grsign=\hbox{$>$} \newdimen\grdimen \grdimen=\ht\grsign
\newbox\simlessbox \newbox\simgreatbox \newbox\simpropbox
\setbox\simgreatbox=\hbox{\raise.5ex\hbox{$>$}\llap
     {\lower.5ex\hbox{$\sim$}}}\ht1=\grdimen\dp1=0pt
\setbox\simlessbox=\hbox{\raise.5ex\hbox{$<$}\llap
     {\lower.5ex\hbox{$\sim$}}}\ht2=\grdimen\dp2=0pt
\setbox\simpropbox=\hbox{\raise.5ex\hbox{$\propto$}\llap
     {\lower.5ex\hbox{$\sim$}}}\ht2=\grdimen\dp2=0pt
\def\simgt{\mathrel{\copy\simgreatbox}}
\def\simlt{\mathrel{\copy\simlessbox}}

\shorttitle{Spectral peak of gamma-ray bursts}
\shortauthors{A. M. Beloborodov}

\begin{document}

\title{Regulation of the spectral peak in gamma-ray bursts}
\author{Andrei M. Beloborodov}
\affil{Physics Department and Columbia Astrophysics Laboratory, Columbia University, 538 West 120th Street, New York, NY 10027, USA
}

\label{firstpage}
\begin{abstract}
Observations indicate that the peak of gamma-ray burst spectrum forms in 
the opaque region of an ultra-relativistic jet. Recent radiative transfer calculations 
support this picture and show that the spectral peak is inherited from initially
thermal radiation, which is changed by heating into a broad photon distribution
with a high-energy tail.  We discuss the processes that regulate the observed position 
of the spectral peak $\Ep$. The opaque jet has three radial zones:
(1) Planck zone $r<\RP$ where a blackbody spectrum is enforced; this 
zone ends where Thomson optical depth decreases to $\tau\approx 10^5$. 
(2) Wien zone $\RP<r<\RW$ with Kompaneets parameter $y\gg 1$ where 
radiation has a Bose-Einstein spectrum, and (3) Comptonization zone $r>\RW$ 
where the radiation spectrum develops the high-energy tail.
Besides the initial jet temperature, an important factor regulating $\Ep$ is 
internal dissipation (of bulk motions and magnetic energy) at large distances 
from the central engine. Dissipation in the Planck zone reduces $\Ep$, and 
dissipation in the Wien zone increases $\Ep$. In jets with sub-dominant magnetic 
fields, the predicted $\Ep$ varies around 1~MeV up to a maximum value of about 
10~MeV. If the jet carries an energetically important magnetic field, $\Ep$ can be 
additionally increased by dissipation of magnetic energy. This increase is hinted 
by observations, which show $\Ep$ up to about 20~MeV. We also consider 
magnetically dominated jets; then a simple model of magnetic dissipation gives 
$\Ep\approx 30\,\GamW$~keV where $\GamW$ is the jet Lorentz factor
at the Wien radius $\RW$.
\end{abstract}



\section{Introduction}

Observed spectra of gamma-ray bursts (GRBs) peak at energy $\Ep$ that 
varies around 1~MeV (after correcting by $1+z$ for the cosmological redshift,
Kaneko et al. 2006; Goldstein et al. 2012).
The spectrum shape can be described by a simple Band function 
(Band et al. 2009) ---  two power laws that are smoothly connected at $\Ep$.
Bursts of higher luminosity are observed to have higher $\Ep$. An approximate 
correlation $\Ep\approx 0.3\,L_{\gamma,52}^{1/2}$~MeV was reported
(e.g. Wei \& Gao 2003; Yonetoku et al. 2004; Ghirlanda et al. 2011), where 
$L_{\gamma,52}$ is the burst luminosity (isotropic equivalent) in units of 
$10^{52}$~erg~s$^{-1}$.
 
The present paper addresses the origin of the spectral peak and the processes
that regulate its position.

\subsection{Synchrotron model}

A simple phenomenological GRB model posits that we observe synchrotron 
radiation, in analogy with blazar models. The model assumes that a nonthermal 
electron population is injected in the jet by some dissipative process. It
gives a spectrum with 
\beq
\label{eq:synch}
  \Ep\approx E_s= 0.4 \Gamma\,\gampeak^2\,\hbar\,\frac{eB}{m_ec}.
\eeq
Here $B$ is the magnetic field measured in the rest frame of the jet (``fluid frame''),
$\gampeak$ is the Lorentz factor at which the injected electron distribution 
peaks (also measured in the fluid frame), and $\Gamma$ is the bulk Lorentz 
factor of the jet itself. If the injection distribution above $\gampeak$ is a 
power-law $dN_e/d\gamma\propto \gamma^{-p}$ then 
the synchrotron spectrum has a high-energy power-law tail,
$dN_\gamma/dE\propto E^{-p/2-1}$ at $E>\Ep$ .

One possibility for the injection of high-energy electrons is associated with 
internal shocks (Rees \& M\'esz\'aros 1994). 
A mildly relativistic electron-ion shock produces an electron distribution with
$\gampeak=\ep_e(m_p/m_e)$, where $\ep_e$ can be a significant fraction
of unity. This gives
\beq
\label{eq:synch1}
    \Ep\approx 1\,r_{12}^{-1}\ep_B^{1/2} L_{52}^{1/2}\left(\frac{\ep_e}{0.3}\right)^2
            {\rm ~MeV}, 
\eeq
where $r_{12}$ is radius in units in $10^{12}$~cm, $L_{52}$ is the isotropic 
equivalent of the jet power in units of $10^{52}$~erg~s$^{-1}$, and $\ep_B$
is the fraction of jet energy that is carried by the magnetic field. If the shock 
heating radius happens to be $r\sim 10^{13}\,\ep_B^{1/2}\ep_e^2$~cm then 
$\Ep$ would be consistent with observations. 

Significant, perhaps dominant, magnetic fields are expected in GRB jets. 
The field is advected by the conducting 
plasma from the central engine, and plausible scenarios (e.g. 
hyper-accretion disks around black holes or proto-neutron stars) invoke 
strong fields. At radii much larger than the size of the central engine 
the advected field is transverse to the jet direction. 
Simulations of shocks in the presence of transverse magnetic field with 
$\ep_B>10^{-4}-10^{-3}$ show no particle acceleration to high energies
$\gamma\gg\gampeak$ (Sironi \& Spitkovsky 2011). Thus, synchrotron emission 
from shocks is not expected to extend far above $\Ep$, which conflicts with 
observations.

The problem with electron acceleration in internal shocks may be avoided if
the synchrotron model is viewed more broadly as a phenomenological
model that does not specify the origin of nonthermal electrons.
When tested against data, the model encounters the following problems.

(1) Thousands of GRBs have been observed, and most of them have $\Ep$
near 1~MeV (Goldstein et al. 2012). Few bursts have $\Ep$ above 10~MeV 
and no bursts are known with $\Ep>20$~MeV.
The synchrotron model does not explain the clustering of $\Ep$
around 1~MeV. The model predicts $\Ep\propto \Gamma\gampeak^2 B$ which 
should give a broad distribution --- there is no reason for this 
combination of $B$, $\gampeak$, and $\Gamma$ to be comparable in 
different bursts, or even within one burst, as GRBs are strongly variable. 

(2) High-energy electrons quickly cool to $\gamma<\gampeak$ (which makes
the process radiatively efficient) and should emit radiation at $E<\Ep$ with 
photon index $\alpha=-3/2$. A typical low-energy index observed
in GRBs is $\alpha= -1$, and many bursts have even harder slopes $\alpha>0$
(Kaneko et al. 2006).
The observed hard slopes are in conflict with the synchrotron model.

(3) The observed spectral peak is sharp. $\Ep$ is defined as photon energy 
at which the burst luminosity peaks, i.e. where $EL_E$ is maximum ($L_E$ 
is the spectral luminosity), and the spectrum shape around the maximum 
can be quantified by its width at half maximum, $E_1<E<E_2$. The observed 
width $\log(E_1/E_2)$ is typically 1-1.5 decades in photon energy. 
The synchrotron model predicts a broader peak (see e.g. the predicted spectra in 
Daigne et al. 2011). To make the spectral peak as sharp as possible, one has 
to assume an unrealistic electron distribution that has a step-like cutoff at
$\gamma<\gampeak$ (Baring \& Braby 2004; Burgess et al. 2011). 
It is not reasonable to expect a step-like electron distribution for a few 
reasons. First, no known acceleration process gives the electron distribution 
with a low-energy cutoff. Second, a low-energy wing of the distribution must be 
created by the fast electron cooling. Third, many GRBs are highly variable, and 
the expected variability in $\gampeak$ should smear out the cutoff in the 
time-averaged emission.

Note also that synchrotron spectra with sharp peaks (that could be fitted 
by a Band function) are not observed in any other astrophysical objects. 
A close example is provided by blazar spectra (e.g. Ghisellini 2006). 
Their synchrotron spectra have the half-maximum width of several 
orders of magnitude, much broader than in GRBs.

The problems of the synchrotron model are shared by other versions of 
optically thin emission, e.g. jitter radiation. Observations 
suggest that the GRB spectral peak forms in the opaque region of the jet.

\subsection{Photospheric emission}

In the opaque jet, photons keep interacting with the plasma and
their spectrum is expected to take a well defined shape with a sharp peak. 
(For example, consider the extreme case of a Planck distribution.)
The radiation is released near the photospheric radius $\Rph$, and
a distant observer will see a spectrum with a sharp peak.

The simplest model of photospheric emission assumes a freely expanding 
radiation-dominated outflow with no baryon loading or magnetic field 
(Paczy\'nski 1986; Goodman 1986). It was recently shown that the emission 
received by distant observers from such outflows has a Planck spectrum 
(Beloborodov 2011). The peak energy of the Planck spectrum is related to
the average photon energy $\Eav$ by $\Ep\approx 1.45\Eav$.
In the ideal radiation-dominated outflow $\Eav$ remains constant, equal
to its value near the central engine $\Ein$. 

In general, $\Ein$ may be expressed in terms of the jet power $L_0$
and the initial radius $r_0$ (comparable to the size of the central compact object),
\beq
\label{eq:E0}
   \Ein\approx 10\, \ep_0^{1/4}\,L_{0,52}^{1/4}\, r_{0,6}^{-1/2}~{\rm MeV},
\eeq
where $\ep_0$ is the initial thermal fraction of the power $L_0$.
Radiation-dominated jets have $\ep_0=1$.
Their predicted $\Ep\approx 1.45\Ein$ may be made consistent with observed 
$\Ep\sim 0.3\,L_{\gamma,52}^{1/2}$~MeV if $r_0$ is large and the flow is 
collimated within a small angle $\theta_b$, which reduces the true jet power, 
$L_0\approx(\thb^2/2) L_\gamma$.

This simple model, however, fails to explain the observed spectra.
Although the Planck spectrum may appear in the time-resolved emission of some 
bursts (e.g. Ryde et al. 2011), GRB spectra are typically nonthermal, with 
an extended high-energy tail.

Theoretically, GRB jets may be expected to carry baryons and magnetic fields, 
and this more detailed model offers an explanation of the observed spectra. 
Two types of jets may be considered:

(1) {\it Thermally dominated baryon-loaded jets:} at small radii the jet is dominated by the thermal energy of radiation (and $e^\pm$ pairs).
The expanding fluid cools adiabatically and its thermal energy is converted to 
the bulk kinetic energy of the baryonic flow (Paczy\'nski 1990; Shemi \& Piran 
1990). Any subphotospheric heating is expected to change the spectrum emitted 
at the photosphere (e.g. Eichler \& Levinson 2000; Rees \& M\'esz\'aros 2005; 
Pe'er et al. 2006). In particular, collisional dissipation was shown to peak at 
Thomson optical depths $\tau\sim 10$ and its detailed calculations yielded 
spectra consistent with observations (Beloborodov 2010; Vurm et al. 2011).
The calculations show that synchrotron emission significantly contributes
to the photospheric emission at $E<\Ep$ but never dominates the spectral peak.
The peak is dominated by radiation that has been thermalized at radii $r\ll\Rph$.

(2) {\it Magnetically dominated jets:} at small radii the jet energy in the fluid frame
is dominated by the magnetic field. 
In the lab frame, the jet luminosity is dominated by the Poynting flux.
The magnetic field gradually dissipates and the jet Lorentz factor grows
with radius. For instance, Drenkhahn \& Spruit (2002) 
considered an alternating magnetic field that dissipates via reconnection. 
The jet also carries baryons, and the ultimate result of dissipation can be the 
conversion of magnetic energy to the bulk kinetic energy of baryons, with some
radiative losses. This gradual conversion can take several decades in radius; 
it peaks at a radius $\Rdiss$ which may or may not be comparable to the 
photospheric radius $\Rph$.
Another uncertainty is the unknown effect of magnetic dissipation on the electron 
distribution. Assuming Maxwellian electrons, Giannios (2008) calculated 
the radiation produced by magnetic dissipation. The predicted spectra 
are in reasonable agreement with observed GRBs if $\Rdiss$ is comparable to $\Rph$.

Radiative transfer in a heated subphotospheric region has been studied with 
four different numerical codes (Pe'er et al. 2006; Giannios 2008; Beloborodov 2010;
Vurm et al. 2011), consistently giving Band-type spectra. These calculations 
show how the spectrum broadens from the thermal to Band shape
at optical depths $\tau\simlt 30$. The resulting $\Ep$ at $\tau\sim 1$ is 
weakly changed from its value at $\tau\sim 30$ (Beloborodov 2010; 
Giannios 2012).\footnote{ 
  Radiative transfer calculations show that, without 
  subphotopsheric heating, adiabatic cooling would reduce $\Ep$ by a factor of 
  $2\tau^{-2/3}$ (Beloborodov 2011). Heating offsets this effect. 
  In the Comptonization zone $\tau<10^2$, the subphotospheric heating 
  is also spent to create the high-energy tail of the spectrum, which limits the 
  growth of $\Ep$. As a result, $\Ep$ remains roughly constant 
  in the Comptonization zone.
  }
Thus, $\Ep$ is mainly regulated at smaller radii where $\tau\gg 30$.  We focus 
in this paper on radiative processes that occur in this highly opaque zone.

\subsection{Photon production and $\Ep$}

The processes regulating the peak of the photospheric spectrum are 
more sophisticated than assumed in existing data analysis. The simplest
estimate $\Ep\approx 4\Gamma\,k\Teff$ associates $\Ep$ with the 
effective temperature of the photosphere $\Teff$, which is defined by
\beq
\label{eq:Teff}
   \frac{4}{3}\,a\Teff^4\,\Gamma^2\,4\pi\Rph^2 \,c=L_\gamma.
\eeq
The estimate $\Ep\approx 4\Gamma\,k\Teff$ simply posits a blackbody 
photospheric emission. In fact, it should be viewed as a lower limit for $\Ep$, 
not its true value.

An improved estimate assumes that radiation is blackbody at $r\sim\RW$ 
instead of $r\sim\Rph$ (Giannios 2012). This assumption can still significantly 
underestimate $\Ep$.
The more realistic model should not make blackbody assumptions at any radii. 
Instead, it must address the number of photons produced in the jet. 
This number is typically not sufficient to maintain a blackbody spectrum with 
$T=\Teff$. The same luminosity $\Lrad$ carried by a smaller number of 
photons implies a higher $\Ep$.

The radiative processes that control the photon number in the opaque thermal 
plasma are detailed in \Sect~2. The problem resembles the evolution of radiation 
in the early universe, although the early universe is known to be much less 
dissipative than GRB jets --- the observed cosmic microwave background
has a nearly Planck spectrum and provides stringent upper limits on 
subphotospheric dissipation.
\Sect~2 also briefly discusses the (uncertain) role of
nonthermal processes at high optical depths.
\Sects~3 and 4 consider thermally and magnetically dominated jets, respectively.
The results are discussed in \Sect~5.


\section{Planck and Wien zones}

\subsection{Notation}

We first introduce notation for basic quantities that will be used in this paper.
Let $dL/d\Omega$ be the jet power per unit solid angle; the 
corresponding isotropic equivalent is defined by $L=4\pi\,(dL/d\Omega)$.
The power is carried by photons, baryons, electrons, $e^\pm$ pairs,
and magnetic field.
The baryonic component includes protons and neutrons; 
  the neutron-to-proton ratio 
depends on the details of the central engine (Beloborodov 2003).
An important parameter of the jet is its energy per proton,
\beq
\label{eq:eta}
     \eta=\frac{L}{\dM c^2},
\eeq
where $\dM=4\pi\,(d\dM/d\Omega)$ is the proton 
mass outflow rate (isotropic equivalent). 
The proton number density in the 
rest frame of the jet (``fluid frame'') is given by
\beq
   n=\frac{\dM}{4\pi r^2 m_p c\, \Gamma},
\eeq
where $\Gamma(r)$ is the Lorentz factor of the jet. 
In addition to the proton-electron plasma
the jet may contain $e^\pm$ pairs of density $n_\pm$. 

The expansion timescale measured in the fluid frame is given by
\beq
\label{eq:texp}
    \texp=\frac{r}{c\Gamma},
\eeq
and the characteristic optical depth of the jet at a radius $r$ is 
\beq
\label{eq:tau}
    \tau=\frac{(n+n_\pm)\,\sT r}{\Gamma}.
\eeq
We will focus on the opaque region $\tau\gg 1$.

The radiation component of the jet power (isotropic equivalent) can be expressed as 
\beq
\label{eq:Lrad}
   \Lrad=4\pi r^2 \frac{4}{3}\Urad \Gamma^2 c,
\eeq
where $\Urad$ is the radiation energy density in the fluid frame, and $(4/3)\Urad$ 
is the radiation enthalpy. The fraction of the total jet power that is carried by radiation
is 
\beq
\label{eq:ep}
    \ep=\frac{\Lrad}{L}=\frac{4}{3}\,\frac{\Gamma\,\Urad}{\eta\, n\, m_pc^2}.
\eeq
As the jet expands, $\ep$ may evolve as a result of adiabatic cooling and 
dissipative heating.

The magnetic (Poynting flux) component of the jet power is 
\beq
    L_B=r^2 B^2 \Gamma^2 c.
\eeq
In this paper, the magnetic fraction will be defined by
\beq
   \ep_B=\frac{L_B}{L},
\eeq
i.e. normalized to the total jet power, so that $\ep_B\leq 1$.

\subsection{Planck zone}

The central engines of GRBs are hot and fill their jets with blackbody radiation
of a high temperature. As the jet 
expands, the Planck spectrum is initially enforced by huge rates 
of photon emission and absorption. The number density of Planck photons
at temperature $T$ is given by
\beq
\label{eq:nP}
   \nP\approx \frac{0.2}{\lbar^3}\,\Th^3, \qquad \Th=\frac{kT}{m_ec^2},
\eeq
where $\lbar=\hbar/m_ec$ is Compton wavelength.
The temperature of blackbody radiation is determined by its energy density, 
$aT^4=\Urad$, where $a=\pi^2 k^4/15c^2\hbar^3$.

There is a characteristic ``Planck radius'' $\RP$ inside of which radiative processes 
in the thermal plasma are fast enough to enforce a Planck spectrum.
Outside $\RP$ photon production becomes inefficient and the photon number 
freezes out (until the jet expands to radii where nonthermal processes may 
produce significant synchrotron radiation). Below we show that the jet temperature 
at $\RP$ is $\ThP\approx 0.01-0.02$, almost independent of the GRB parameters.

Two radiative processes should be considered in the thermal plasma: 
bremsstrahlung $e+p\rightarrow e+p+\gamma$ and double Compton effect
$e+\gamma\rightarrow e+\gamma+\gamma$. Let $\dnB$ and $\dnDC$ be 
the rates of photon production by these two processes [cm$^{-3}$~s$^{-1}$].
In a plasma with approximately Planckian radiation the rates are given by
\beq
\label{eq:dnB}
   \dnB=\xi\, n^2\sT c\,\Th^{-1/2},
\eeq
\beq
\label{eq:dnDC}
   \dnDC=\chi\, n_\gamma n\,\sT c\, \Th^2,
\eeq
where the numerical factors $\xi\approx 0.06$ and $\chi\approx 0.1$ 
weakly (logarithmically) depend on $\Th$ (see Appendix).
When evaluating the rates of photon production we neglected the 
presence of $e^\pm$ pairs; this approximation is reasonable as will be 
discussed below. The total rate of photon production is 
$\dn_\gamma=\dnDC+\dnB$.

Ratio $\dnDC/\dnB\approx 2(n_\gamma/n)\Th^{5/2}$ is sensitive
to temperature and also depends on the photon-to-baryon ratio, 
which may be evaluated as 
\beq
\frac{n_\gamma}{n}=\frac{\Lrad}{\Eav} \frac{m_p}{\dM}
                               = \frac{\ep\,\eta\,m_pc^2}{\Eav}
  = 3\times 10^5 \ep\,\left(\frac{\eta}{300}\right) \left(\frac{\Eav}{\rm MeV}\right)^{-1}.
\eeq
Uisng $n_\gamma/n\sim 10^5$, one concludes that $\dnDC>\dnB$ 
for $\Th\simgt 0.01$.\footnote{
    This conclusion may not hold 
    if the jet is polluted with clumps or sheets with a low entropy per baryon. 
    In such clumps bremsstrahlung could dominate photon production.}
For the estimates of the boundary of the Planck zone 
we will use the approximation $\dn_\gamma\approx \dnDC$.
 
The balance between emission and absorption of Planck photons is maintained 
as long as $\dn_\gamma\texp>n_\gamma$, and radius $\RP$ is defined by the 
following condition,
\beq
\label{eq:RP}
     \texp \dnDC=n_\gamma.
\eeq
Substituting \Eqs~(\ref{eq:texp}) and (\ref{eq:dnDC}), and using 
\Eq~(\ref{eq:tau}) with $n_\pm\ll n$, we can rewrite the condition (\ref{eq:RP}) as
\beq
\label{eq:RP1}
      \Th^2\tau=\chi^{-1}.
\eeq
As radiation is still approximately blackbody at $r\sim\RP$, we can use 
$aT^4\approx\Urad$ and substitute $\Urad$ from \Eq~(\ref{eq:ep}). 
This gives,
\beq
\label{eq:TBB}
    T\approx\Teff=\left(\frac{3\,\ep\,\eta\,n\,m_pc^2}{4\,\Gamma a}\right)^{1/4}
      =\left(\frac{3\,\ep\,\eta\,m_pc^2\tau}{4\,a\,\sT r }\right)^{1/4}.
\eeq
From \Eqs~(\ref{eq:RP1}) and (\ref{eq:TBB}), we find $\Th$ at the Planck radius,
\begin{eqnarray}
\label{eq:ThP}
\nonumber
   \ThP & = &\left(\frac{45}{4\pi^2\chi} \frac{m_p}{m_e}\,
                     \epP\,\eta\,\frac{\lbar^3}{\sT\RP}\right)^{1/6}  \\
           & \approx & 1.1\times 10^{-2}\,R_{{\rm P},10}^{-1/6}\epP^{1/6}\eta_{2}^{1/6}.
\end{eqnarray}
The scattering optical depth at the Planck radius is 
\beq
\label{eq:tauP}
  \tauP=(\chi\ThP^2)^{-1}\sim 10^5. 
\eeq
Note that $\ThP$ and $\tauP$ weakly depend on the exact value of $\RP$, 
which  can be determined in concrete jet models discussed 
in Sections~3 and 4. 

Our calculation of $\ThP$ neglected $e^\pm$ pairs. This approximation 
is easy to check for the thermal plasma. At $r\simlt\RP$ pair annihilation 
balance is maintained with approximately blackbody radiation field. 
At $\Th\ll1$ it gives the following equation for positron density $n_+$ 
(e.g. Svensson 1984),
\beq
   \frac{n_\gamma^2}{n_+(n_++n)}=\frac{8\,\zeta^2(3)}{\pi}\,\Th^3
                                                        \exp\left(\frac{2}{\Th}\right),
\eeq
where $\zeta(3)\approx 1.2$ is the Riemann zeta function.
One can see that $n_+\ll n$ at $\Th=\ThP$.

 \subsection{Wien zone}

Compton scattering maintains a Bose-Einstein distribution of photons
as long as Kompaneets parameter 
\beq
\label{eq:y}
   y=4\Th\tau\gg 1.
\eeq 
Comptonization with $y\gg 1$ enforces a common electron-photon 
temperature, $T_e=T_\gamma=T$.
The strong thermal coupling between photons and electrons may also 
be viewed from the electron point of view.
The condition $y\gg 1$ is equivalent to  $(3/2)nkT/\tC\gg\Urad/\texp$
where $\tC=3m_ec/8\sT\Urad$ is the timescale for electron temperature 
relaxation to the radiation temperature. 

The Bose-Einstein distribution is described by the photon occupation number
\beq
\label{eq:Wien}
     {\cal N}=\frac{1}{\exp(\mu+x)-1}, \qquad x=\frac{h\nu}{kT},
\eeq 
where chemical potential $\mu\geq 0$ describes the deficit of photon
density compared with the Planck value $\nP$.
In the Planck zone, the distribution takes the Planck form ($\mu\ll 1$) 
with $n_\gamma=\nP$. Outside the Planck radius the photon number 
freezes out and, in a heated jet, $n_\gamma/\nP<1$.
Thermalized radiation with $n_\gamma\ll\nP$ is described by $\mu\gg 1$; 
then radiation has a Wien spectrum
${\cal N}\approx \exp(-\mu-x)$.

Hereafter we call the region where $y>1$ the ``Wien zone'' 
to emphasize the possibility of $\mu\gg 1$ outside $\RP$.
More exactly, photons must have a Bose-Einstein distribution where 
$y\gg 1$; whether $\mu\gg 1$ is satisfied depends on the heating history. 
The average photon energy in the fluid frame is between $2.7kT$ 
(Planck) and $3kT$ (Wien). 
Radiation energy density in the Wien zone is given by 
\beq
\label{eq:UW}
    \Urad\approx 3kT\,n_\gamma,  \qquad   r<\RW.
\eeq 
As the photon density $n_\gamma$ may be below $\nP=aT^4/2.7kT$, 
the radiation density $\Urad$ may be below the blackbody value $aT^4$.

\Eqs~(\ref{eq:ep}) and (\ref{eq:UW}) give the following expression
for the photon-to-baryon ratio,
\beq
\label{eq:ratio}
    \frac{n_\gamma}{n}=\frac{\eta\, m_p}{4m_e}\,\frac{\ep}{\Gamma\Th}.
\eeq
The average photon energy in the fixed lab frame (also the frame of a distant 
observer) is given by 
\beq
\label{eq:Eav}
    \Eav=\frac{\ep\,\eta\,m_pc^2}{n_\gamma/n}=4\Gamma kT.
\eeq
The thermal (Bose-Einstein) spectrum peaks at a photon energy that is 
slightly above $\Eav$,
\beq
   \Ep=\Eav \times \left\{ \begin{array}{cc}
                                      1.45  &  {\rm Planck} \\
       \displaystyle{\frac{4}{3}}  &  {\rm Wien}
                    \end{array} \right.
\eeq

The observed photospheric spectrum is changed from the Wien shape
by Comptonization outside the Wien zone; however, $\Ep$ weakly evolves 
outside $\RW$. Thus, the observed $\Ep$ may be estimated as $\EavW$, 
which is given by
\beq
\label{eq:EW}
      \EavW\approx \frac{\epW}{\epP}\,\EavP
            \approx \frac{\epW}{\epP}\,4\GamP k\TP.
 \eeq
Here index ``P'' indicates that the quantity is evaluated at the Planck radius
and index ``W'' --- at the Wien radius. Equation~(\ref{eq:EW}) assumes that the photon 
number carried by the jet is not significantly changed outside the Planck zone;
then $\Eav\propto \ep$ between $\RP$ and $\RW$. 
\Sect~2.2 showed that $k\TP=5-10$~keV, as long as photon production is 
dominated by the thermal plasma. Then,
\beq
\label{eq:Ep1}
    \Ep\approx  30\,\GamP\,\frac{\epW}{\epP} {\rm ~keV}.
\eeq

The optical depth at the Wien radius $\RW$ may be expressed in terms of 
$\ThW$ from $y\approx 1$,
\beq
\label{eq:tauW}
   \tauW\approx \frac{1}{4\ThW}\approx \frac{\GamW \, m_ec^2}{\Ep}.
\eeq
Maintaining the Bose-Einstein spectrum requires $y\gg 1$, which corresponds to 
$\tau\gg\tauW$.

\subsection{Evolution equation for photon-to-baryon ratio}

The transition between the Planck and Wien zones is 
more accurately described by an evolution equation for $n_\gamma/n$. 
Using $dt=dr/\Gamma c$ (where $t$ is the proper time measured in the fluid frame)
the rate of change of $n_\gamma/n$ may be written as
\beq
\label{eq:diff1}
   \frac{d}{dr}\left(\frac{n_\gamma}{n}\right)=\frac{\dn_\gamma}{n\,\Gamma\,c},
\eeq
where $\dn_\gamma$ is the net rate of photon production measured in the 
fluid frame,
\beq
\label{eq:dng}
   \dn_\gamma=\left(\dnB+\dnDC\right)\left(1-\frac{n_\gamma}{\nP}\right).
\eeq
Here $\nP(\Th)$ is the Planck density (\Eq~\ref{eq:nP}), and the factor 
$(1-n_\gamma/\nP)$ takes into account photon absorption; the net photon 
production rate vanishes when $n_\gamma=\nP$ as expected from detailed
balance for blackbody radiation. 

To obtain the equation for function $f(r)=n_\gamma/n$ first note that
temperature is related to $f$ by \Eq~(\ref{eq:ratio}), which we rewrite as
\beq
    \Th=\frac{H}{f}, \qquad H(r)=\frac{\eta\, m_p}{4 m_e}\,\frac{\ep(r)}{\Gamma(r)}.
\eeq
Then from \Eq~(\ref{eq:diff1}) we obtain,
\beq
\label{eq:diff}
    \frac{df}{d\ln r}=\tau\,\left(\xi\,\frac{f^{1/2}}{H^{1/2}}+\chi\,\frac{H^2}{f}\right)
                        \left(1-\frac{\lbar^3 n\,f^4}{0.2\,H^3}\right).
\eeq
This equation is easily solved for $f(r)$ for any concrete jet model with given
$\Gamma(r)$, $n(r)$, and heating history $\ep(r)$. 
The solution also determines $\Th(r)=H/f$.

\subsection{Nonthermal processes}

We focus in this paper on the early, opaque stages of expansion and consider 
thermal heating of the jet due to dissipation of internal bulk motions or magnetic
energy. It is, however, possible that dissipation also generates high-energy
nonthermal particles, even at very high optical depths. This possibility is  
questionable for internal shocks --- it was argued that shocks at high optical 
depths are mediated by radiation and have a considerable thickness, comparable 
to the photon free path (Levinson 2012); such shocks would be unable to 
accelerate electrons. Nonthermal electrons can be generated by magnetic 
reconnection, although the efficiency of this process is uncertain.
High-energy electrons produce synchrotron photons that can be Comptonized 
to the Wien peak and contribute to $n_\gamma$ (Thompson et al. 2007). 

Here we limit our discussion to the following estimate.
Suppose a fraction $\epnth$ of the jet power $L$ is given to accelerated
electrons. They immediately radiate this power via inverse Compton scattering
and synchrotron losses. Scattering does not change photon number, so 
only the synchrotron luminosity is relevant, which is given by
\beq
  L_s=\frac{U_B}{\fKN\Urad+U_B}\,\epnth\,L
        =\frac{\ep_B/2}{\fKN(3/4)\ep+\ep_B/2}\,\epnth\,L.
\eeq
Here $U_B=B^2/8\pi$, $\ep_B$ and $\ep$ are the fractions of jet energy carried 
by the magnetic field and radiation, respectively; the factor $\fKN<1$ describes 
the Klein-Nishina correction to Compton losses.

The synchrotron power $L_s$ peaks at energy $E_s$ given by \Eq~(\ref{eq:synch}),
assuming $\gampeak$ is high enough to avoid synchrotron self-absorption.
The produced photon {\it number}, however, peaks at the low end of the synchrotron
spectrum. The photon production may be roughly estimated as\footnote{
     The power index of $(E_s/\Eabs)^{1/2}$ in this equation comes from the
     standard synchrotron spectrum of fast-cooling electrons; it may be slightly 
     changed if the 
     injected electron distribution is modified by $e^\pm$ cascade  
     accompanying inverse Compton scattering.}
\beq
  \dot{N}_s\approx \frac{L_s}{E_s}\,\left(\frac{E_s}{\Eabs}\right)^{1/2},
\eeq
where $\Eabs$ is the energy above which soft photons are Comptonized
to the Wien peak faster than absorbed. 
Photons may be 
self-absorbed by the high-energy electrons or absorbed by the thermal plasma
via inverse double Compton effect (Appendix); typically $\Eabs\sim 10^{-2}\Eav$. 
More careful calculations take into account the induced downscattering (Bose
condensation) of the synchrotron photons on the thermal electrons, which 
increases the effective $\Eabs$ (Vurm et al. 2012). 

Comparing $\dot{N}_s$ with the existing flux of thermal photons, one finds
\beq
    \frac{\dot{N}_s}{\dN}\approx \frac{\epnth\ep_B}
    {\ep\left(\frac{3}{2}\,\fKN\ep+\ep_B\right)}\,\frac{\Eav}{E_s}
        \,\left(\frac{E_s}{\Eabs}\right)^{1/2}.
\eeq
Synchrotron emission does not appreciably change the photon number carried 
by the jet when this ratio is smaller than unity. The highest $\dot{N}_s$ is 
achieved if $E_s\sim\Eabs$. Even this maximum rate may be insufficient to 
significantly influence the photon number  at large optical depths; it depends 
on the unknown $\epnth$ and the Lorentz factors of the accelerated particles.


\section{Thermally dominated jets}

\subsection{Non-dissipative jet}

Early works on GRBs studied in detail the dynamics of ideal (non-dissipative) 
relativistic hot outflows loaded with baryons
(e.g. Paczy\'nski 1990; Shemi \& Piran 1990).
The flow acceleration is controlled by parameter $\eta$ (\Eq~\ref{eq:eta}).
In a radial flow, the fluid Lorentz factor $\Gamma$ grows linearly with $r$ until 
it approaches its maximum value $\Gamma=\eta$ at $\Rsat\sim \eta r_0$. 

The photospheric radius $\Rph$ is defined by $\tau=1$; it is given by
\beq
\label{eq:Rph}
    \Rph=\frac{L \sT (1+n_\pm/n)}{4\pi\, m_p c^3\, \Gamma^2\eta}.
\eeq
We will consider here jets with $\Rph>\Rsat$, so that $\Gamma\approx\eta$ at 
the photosphere. In the absence of dissipation, the density of relict $e^\pm$ 
pairs is negligible at $\Rph$, $n_\pm/n\ll 1$.

In the non-dissipative jet, the evolution of radiation is fully controlled by 
adiabatic cooling. Entropy is dominated by radiation and proportional to the 
photon number; thus, adiabatic cooling conserves photon number. 
The photon-to-baryon ratio $n_\gamma/n$ remains constant as the jet 
expands;\footnote{More exactly, the photon-to-baryon ratio is 
      $n_\gamma/(n+n_n)$
      where $n_n$ is the number density of neutrons. For simplicity, 
      we assume a constant proton fraction $Y_e=n/(n+n_n)$ and 
      use $n_\gamma/n$ as a conserved ``photon-to-baryon ratio.''
      }
it is set by the initial conditions at radius $r_0$,
\beq
\label{eq:ratio1}
   \frac{n_\gamma}{n}\approx 240\, \eta\, r_{0,7}^{1/2}L_{0,52}^{-1/4}.
\eeq  

At radii $r<\Rsat$, where the jet power is dominated by radiation,
$L_\gamma\approx L$, the constancy of the total fluxes of energy and 
photon number implies a constant energy per photon $\Eav(r)=\Lrad/\dN=const$. 
Adiabatic cooling of photons in the fluid frame is compensated by the 
increasing Doppler shift as the jet accelerates, $\Eav\propto T\Gamma=const$.

At radii $\Rsat<r<\Rph$, radiation continues to cool adiabatically, 
$T\propto n^{1/3}\propto r^{-2/3}$ while $\Gamma\approx const=\eta$.
Its spectrum is still blackbody, the photon number is still conserved, and 
$\Lrad$ is decreasing, $\Lrad\propto T\propto r^{-2/3}$.
As a result, $\Eav$ and $\Lrad$ are reduced between $\Rsat$ and $\Rph$ by a 
factor of $\sim(\Rph/\Rsat)^{-2/3}$. This gives
\beq
\label{eq:E1}
     \Eav(\Rph)\approx 4\, \eta_3^{8/3} L_{52}^{-5/12} r_{0,7}^{1/6} {\rm ~MeV}.
\eeq
This standard estimate (e.g. Paczynski 1990) is refined by a factor of two by 
accurate radiative transfer calculations (Beloborodov 2011). The predicted 
photospheric emission $L_\star\approx (\Eav_\star/\Ein)\,L$ is bright and 
has a high $\Ep>1$~MeV if the jet has $\eta\simgt 600 L_{52}^{5/32}$.
Its spectrum cuts off exponentially above $\Eav(\Rph)$.

\medskip

\subsection{Dissipative jet}

Dissipation can offset adiabatic cooling between $\Rsat$ and $\Rph$ and keep 
$\Lrad$ close to the total jet power $L$. Heating is especially important for bursts 
with $\Rph\gg\Rsat$ where adiabatic cooling threatens to greatly reduce $\Ep$.

Deep subphotospheric heating is expected, in particular if $\Rph\gg\Rsat$, 
because such jets have moderate Lorentz factors $\Gamma$ and thus 
internal dissipation should start early. For instance, internal shocks 
can form and propagate at all radii $r>\Gamma^2\lambda$,
where $\lambda$ is the minimum scale of the Lorentz factor variations,
possibly comparable to the size of the central compact object $\sim 10^6$~cm.
The energy density in the shock is dominated by radiation and protons. 
In the region of large optical depth $\tau\gg 10$ the proton heat is quickly shared
with electrons via Coulomb collisions (Beloborodov 2010), and 
Compton scattering immediately passes the heat to radiation, 
which dominates the heat capacity of the jet by a huge factor $\sim n_\gamma/n$.

Similar electron/photon heating is expected in the presence of any mechanism 
that stirs protons and gives them random motions in the fluid frame. It may result 
from magnetic reconnection. Dissipation of subdominant magnetic field 
$B^2/8\pi\simlt nm_pc^2$ is sufficient to give thermal energy $\simlt \Ein$ per 
photon and keep $\Eav$ from falling far below $\Ein$.

Coulomb electron heating continues at $r>\RW$ where it leads to
$T_e>\Trad$ and Comptonization of Wien radiation into a Band spectrum.
Coulomb heating is a two-body process and its efficiency decreases 
proportionally to optical depth $\tau$ as the jet approaches the photosphere.
The resulting photospheric spectrum is shaped by heating and 
transfer effects at optical depths $\tau\sim 10$. 

Besides the thermal Coulomb heating, the spectrum is affected by electrons that 
are injected with energies $\gamma m_ec^2\sim m_\pi c^2\approx 140$~MeV by 
$n$-$p$ or $p$-$p$ collisions. 
These electrons produce additional high-energy photons via inverse Compton 
scattering and low-energy photons via synchrotron emission. Detailed calculations 
of these processes and their effects on the photospheric spectrum are found in 
Beloborodov (2010) and Vurm et al. (2011).

The peak energy of the emerging spectrum is inherited from the thermal plasma
at large optical depths $\sim\tauW\sim 10^2$.\footnote{$\Ep$ is close to but  
      slightly different from $\Eav$, 
      depending on the spectrum shape. In particular, $\Ep\approx 1.45\Eav$ for a 
      Planck spectrum and $\Ep=(4/3)\Eav$ for a Wien spectrum. The main effect 
      of Comptonization at $r>\RW$ is to create a high-energy tail above $\Ep$, and
      it also slightly affects $\Ep$ itself. Here we neglect the shift of $\Ep$
      between $\RW$ and $\Rph$. 
      }
It depends on the photon-to-baryon ratio $n_\gamma/n$, which can freeze out at 
the Planck radius $\RP$ as discussed in \Sect~2. A high radiative efficiency 
$\ep\sim 1$ is expected in the dissipative jet, and then \Eq~(\ref{eq:Ep1}) gives
\beq
\label{eq:EpP}
    \Ep\approx 30\,\GamP {\rm ~keV}.
\eeq
i.e. $\Ep$ is determined by the jet Lorentz factor at the Planck radius.
A typical GRB with $\Ep\sim 1$~MeV has $\GamP\sim 30$; the 
highest observed $\Ep$ corresponds to $\GamP\sim 300$. Note that
the photon freeze-out can occur before the jet acceleration is complete, 
as $\GamP$ can be smaller than the final Lorentz factor $\Gamma\approx\eta$.

The maximum possible $\Ep$ corresponds to radiative efficiency $\ep=1$
and the minimum photon number $n_\gamma/n$. Since dissipation can only 
increase $n_\gamma/n$ from its central value  $(n_\gamma/n)_0$, the 
maximum $\Ep$ is achieved if 
$n_\gamma/n=(n_\gamma/n)_0$. It corresponds to
\beq   
     \EavW^{\max}=\Ein,
\eeq
where $\Ein$ is given by \Eq~(\ref{eq:E0}). The maximum $\Ep\sim \Ein$ 
is achieved if the flow is adiabatic in the Planck zone and dissipative in 
the Wien zone (maintaining $\ep\sim 1$).
Then the expansion at $r<\RP$ gives $\GamP\approx T_0/\TP\approx \Ein/3k\TP$.

If the Planck zone is not adiabatic (i.e. the flow is dissipative), the photon
number is increased by a factor $Q>1$. At the same time, $\GamP$ is reduced 
by the factor $Q^{-1}$ (so that the jet carries the same luminosity $\Lrad\approx L$ 
and the energy conservation law is satisfied). Then $\Ep$ is also reduced by $Q^{-1}$,
\beq
     \frac{n_\gamma}{n}=Q\,\left(\frac{n_\gamma}{n}\right)_0,  
     \quad \GamP=\frac{\Theta_0}{Q\,\ThP}, 
     \quad \Ep\approx \ep\,\frac{\Ein}{Q}.
\eeq

The Planck radius may be evaluated using \Eq~(\ref{eq:RP1}) and the 
expression for the optical depth,
\beq
\label{eq:RP2}
  \tau_P=\frac{L\sT}{4\pi \RP m_pc^3 \eta \GamP^2}=\frac{1}{\chi \ThP^2}.
\eeq
As long as the main producer of photons in the GRB spectral peak is the 
opaque thermal plasma, we can use $\GamP\approx (\epP/\ep)(\Ep/4 k\TP\ep)$,
where $\Ep$ is the observed peak position. Then \Eq~(\ref{eq:RP2}) gives,
\beq
\label{eq:RP3}
   \RP\approx\frac{4\chi\,\sT\,L}{\pi m_pc^3 \eta}
                     \left(\frac{\epP}{\ep}\,\frac{\Ep}{m_ec^2}\right)^{-2} \ThP^4.
\eeq
Combining with \Eq~(\ref{eq:ThP}) for $\ThP$, we find
\beq
\label{eq:RP4}
   \RP\approx 10^{10}\, \ep^{3/5}\epP^{-4/5} \eta_2^{-1/5} L_{\gamma,52}^{3/5} 
                       \left(\frac{\Ep}{300 \rm ~keV}\right)^{-6/5} 
                                                  {\rm cm}.
\eeq
The observed bursts typically have $\Ep\sim300\,L_{\gamma,52}^{1/2}$~keV
(although some bursts deviate from this relation, e.g. bursts with the highest 
$\Ep$); this is consistent with approximately constant $\RP\sim 10^{10}$~cm.

\subsection{Collimation}

Achromatic breaks in GRB afterglow light curves are often interpreted as 
evidence for jet beaming, with a typical opening angle of 5-10$^{\rm o}$.
Beaming helps explain the extremely high apparent luminosities, 
up to $10^{54}$~erg~s$^{-1}$ in some GRBs. Beaming must be achieved 
through a collimation process. It is expected from the pressure confinement 
of the jet by the progenitor star or by a non-relativistic dense wind from the 
outer regions of the accretion disk around the central object. What effect 
can collimation have on observed $\Ep$?

If collimation is not accompanied by significant dissipation, the expanding
jet can be described as an ideal relativistic flow 
confined by a wall that determines the cross section of the
jet $S(r)=S_0(r/r_0)^\psi$, where $r$ is the radial distance along the jet 
axis and $r_0$ is the size of the central engine.
For instance, a parabolic wall gives $S(r)\propto r$, i.e. $\psi=1$,
and uncollimated (radial) expansion is described by $\psi=2$. 
The opening angle of the jet is determined by $\psi$ and the radius 
$\Rcoll$ where the 
wall ends and free expansion begins (e.g. where the jet escapes the 
progenitor star). Between $r_0$ and $\Rcoll$ the opening angle 
decreases as $\thb\approx (r_0/r)^{1-\psi/2}$. The jet Lorentz factor in the 
collimation funnel grows as $\Gamma\approx (r/r_0)^{\psi/2}$ while its
temperature decreases as $\Gamma^{-1}$ (as required by conservation
laws, see e.g. \Sect~3.1 in Beloborodov 2003).
Note that in the funnel $\Gamma\thb\approx 1$;
it implies a marginal causal contact across the jet.
A typical beaming angle $\thb\sim 0.1$ at $\Rcoll$ corresponds to 
$\Gamma\sim 10\ll \eta$ and temperature $\Th\sim 0.1\Th_0>\ThP$,
i.e. collimation is expected to occur in the Planck zone $r<\RP$.

Dissipationless collimation conserves entropy, and hence does not change
photon-to-baryon ratio $n_\gamma/n$. This implies conservation of the
total photon number carried by the jet. It also implies that collimation
does not change the average photon energy, 
$\Eav$. Beaming boosts the isotropic equivalent of luminosity 
$L_\gamma\approx L$ and the isotropic equivalent of photon flux $\dN$ 
by the same factor $\sim\thb^{-2}=(\Rcoll/r_0)^{2-\psi}$. Their ratio 
$\Eav=\Lrad/\dN$ remains unchanged from its value at $r_0$, $\Eav=\Ein$.
Then photospheric emission has $\Eav=\ep\Ein$ as discussed \Sect~3,
so a radiatively efficient burst with $\ep\sim 1$ basically inherits $\Eav$ 
(and hence $\Ep$) from the central region $r\sim r_0$ 
even though $L$ may be strongly increased by beaming.

Next consider dissipative collimation, for instance collimation accompanied 
by shocks (e.g. Lazzati et al. 2009). 
Dissipation generates entropy and hence increases $n_\gamma/n$.
Thus, the total photon number carried by the jet is increased, $Q>1$, and 
hence $\Eav$ (jet energy per photon) is reduced. 

If there is a relation between $\thb$ and $Q$, it leads to a correlated variation 
of $L$ and $\Ep$ (with $\thb$ being the varying parameter). Note that $\thb$ 
must satisfy $\Gamma\thb \simlt 1$ for a causal contact across the jet.
This condition is marginally satisfied for ideal (non-dissipative) collimation 
and easily satisfied for dissipative collimation, as it reduces $\Gamma$
while increasing the jet internal energy. 

Thompson et al. (2007) considered the possibility that $\thb$ always tends 
to its maximum allowed value $\thb\sim\Gamma^{-1}$. They pointed out 
that this gives $\Eav\propto \thb^{-1}$ and hence $\Ep\propto L^{1/2}$,
similar to the observed trend. Their model encounters, however, two difficulties.
First, it has to invoke large variations in $\thb$, not only from burst to burst 
but also within a single burst (as an extended $\Ep$-$\Lrad$ correlation was 
reported in individual GRBs, e.g. Ghirlanda et al. 2011). Second, the model 
posits that GRBs of various apparent luminosities $L\sim L_0\thb^{-2}$ have 
approximately the same true power $L_0\sim 10^{50}$~erg~s$^{-1}$, which 
implies a central temperature $kT_0\approx 1\,r_{0,6}^{-1/2}$~MeV. 
Then the brightest bursts have the highest 
$\Ep\approx 3kT_0\approx 3 r_{0,6}^{-1/2}$~MeV.
It falls short of the observed highest $\Ep\sim 15$~MeV.

\subsection{Expected range of $\Ep$}

Photospheric emission from dissipative jets is affected by beaming and photon 
production as shown in Figure~1. Beaming increases the apparent luminosity, 
and photon production reduces the observed $\Ep$. Adiabatic cooling needs not 
to be considered, as the dissipative jets maintain $\Lrad\sim L$.

\begin{figure}
\epsscale{1.1}
\plotone{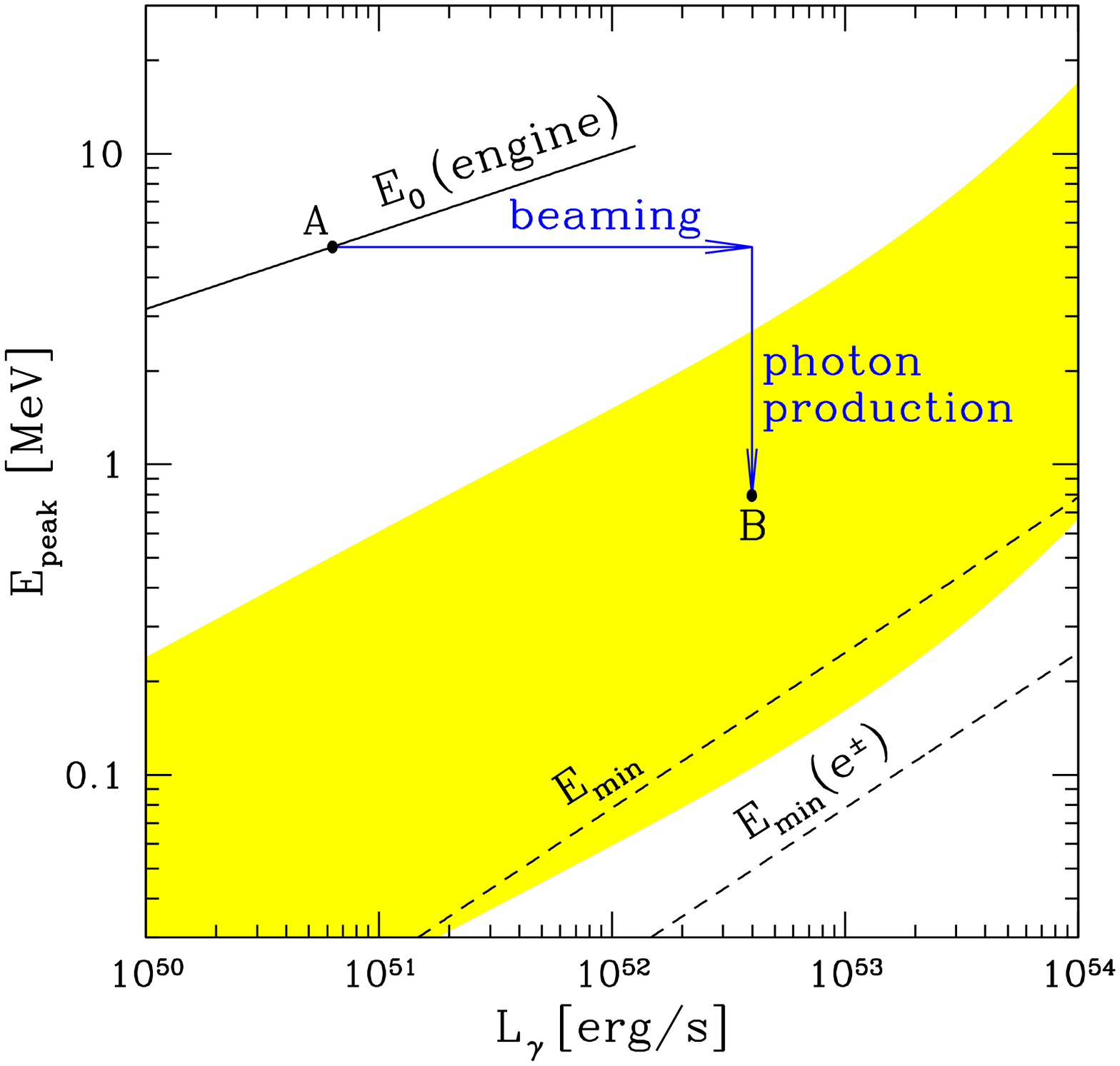}
\caption{
$\Lrad$-$\Ep$ diagram. Point A is an example of the initial condition near the 
central engine; the jet starts with $\Ep$ close to $\Ein$ (\Eq~\ref{eq:E0}).
As the jet expands, its apparent luminosity $\Lrad$ is increased by beaming 
and $\Ep$ is reduced by photon production. Point B shows the resulting 
photospheric emission. The approximate region populated by observed GRBs 
is shown in yellow. The observed $\Ep$ should not violate the lower bound 
$\Emin$ that is set by the effective blackbody temperature of the photospheric 
radiation (\Eq~\ref{eq:Emin}). Dashed line shows $\Emin$ given by 
\Eq~(\ref{eq:Emin1}) with $q=1$. Dashed line marked $\Emin(e^\pm)$ allows 
for the presence of $e^\pm$ pairs with multiplicity $f_\pm=10$. 
}
\end{figure}

Reasonable beaming factors $L/L_0\sim 10^2$ (which are suggested by the burst 
energetics and the afterglow data analysis) together with the expected photon 
production in a dissipative jet naturally explain the location of observed GRBs 
on the $\Lrad$-$\Ep$ diagram (approximately shown by the yellow strip in 
Figure~1). The observed bursts are also consistent with $\RP\sim 10^{10}$~cm
(\Eq~\ref{eq:RP4}). The estimated Lorentz factor at the Planck radius, 
$\GamP\sim 30(\Ep/{\rm MeV})$ (\Eq~\ref{eq:EpP}), is consistent with a slow 
jet acceleration in the zone $r_0<r<\RP$, as expected in the presence of a 
strong collimation.

Accurate theoretical predictions for the burst locations on the diagram are
difficult; however, one can estimate the lower and upper bounds on $\Ep$.
The lower bound is obtained from the fact that the photospheric emission
cannot be colder than the effective blackbody temperature, which is defined 
by \Eq~(\ref{eq:Teff}).
\beq
    \Emin\approx 4 \Gamma k\Teff=2\left(\frac{45}{\pi^3}\right)^{1/4}
                  \left(\frac{\Gamma}{\Rph}\right)^{1/2}\left(\Lrad c^2\hbar^3\right)^{1/4}.
\eeq 
Substitution of \Eq~(\ref{eq:Rph}) for $\Rph$ gives
\begin{eqnarray}
\nonumber
   \Emin & \approx & 4\left(\frac{45}{\pi}\right)^{1/4}
                   \frac{\ep^{1/4}}{f_\pm^{1/2}}\, \eta^{1/2}\Gamma^{3/2}
                   \left(\frac{m_p^2c^8\,\hbar^3}{L\,\sT^2}\right)^{1/4}  \\
              & \approx & 40\,\frac{\eta_2^{1/2}\Gamma_2^{3/2}\ep^{1/4}}
                                               {L_{52}^{1/4}f_\pm^{1/2}} {\rm ~keV},
\label{eq:Emin}
\end{eqnarray}
where $\ep=L_\gamma/L\sim 1$ is the radiative efficiency, and 
$f_\pm=1+n_\pm/n$ 
describes the increase of $\Rph$ due to possible $e^\pm$ creation.

\Eq~(\ref{eq:Emin}) may be simplified if there is a relation between 
$L$ and $\Gamma$ (or $\eta$). The existence of such a relation in GRBs is uncertain.
Model-dependent analysis of afterglow light curves by Ghirlanda et al. (2012) and
L\"u et al. (2012) suggests an approximate correlation $\Gamma\propto L^m$ with 
$m$ between $1/4$ and $1/2$. This (debatable) correlation may motivate one to 
consider a relation of the form $\Gamma_2^2=q\,L_{\gamma,52}^{3/4}$, where 
$q\sim 1$ is a factor that may weakly depend on $\Lrad$. With such 
parametrization, $\Emin$ becomes proportional to $L^{1/2}$,
\beq
\label{eq:Emin1}
   \Emin\approx 40\,q\,L_{52}^{1/2}  f_\pm^{-1/2} {\rm ~keV},
\eeq 
where $\ep\sim 1$ and $\Gamma\approx\eta$ have been assumed 
at the photosphere.

The upper bound on $\Ep$ for thermally dominated jets is set by the initial 
conditions near the central engine. The spectrum of radiation carried by the 
jet peaks at photon energy slightly above $\Eav$, e.g. $\Ep=(4/3)\Eav$ for 
a Wien spectrum. The highest $\Eav$ is set by the jet energy per photon 
near the central engine,
\beq
\label{eq:Epmax}
   \Ein\approx 2.7kT_0, \qquad aT_0^4=U_0,
\eeq
where $U_0$ is the energy density at the base of the 
jet.\footnote{A more accurate expression for $U_0$ includes the 
       contribution from $e^\pm$ pairs. The corresponding correction to 
       $T_0$ (the reduction by a factor of $[4/11]^{1/4}$) is compensated by 
       additional photon energy that appears  in the expanding jet when 
       the $e^\pm$ pairs cool down and annihilate. 
       Therefore, one can use the simplified \Eq~(\ref{eq:Epmax}) for the estimate 
       of the maximum $\Eav$ at radii $r\gg r_0$.  
        }
The energy density is related to the jet power $L_0$ by
\beq
    L_0=\Omega_0\, r_0^2\, U_0 c\beta_0,
\eeq
where $\Omega_0$ is the opening solid angle of the flow near the central
engine, and $\beta_0=v_0/c\sim 1$ is the flow velocity. This gives
\beq
\label{eq:Eav0}
   \Ein\approx 10\,\left(\beta_0\,\frac{\Omega_0}{4\pi}\right)^{-1/4}
                               L_{0,52}^{1/4}\,r_{0,6}^{-1/2} {\rm ~MeV}.
\eeq

If the central engine is an accreting black hole, $r_0$ is comparable to a 
few Schwarzschild radii $r_g=2GM/c^2$, and the maximum possible power 
$L_0$ is comparable to the accretion power, 
\beq
   \Lacc\sim\frac{GM\dMacc}{r_0}\approx  3\times 10^{53}
        \,\dm\,
        \left(\frac{r_0}{3\,r_g}\right)^{-1} {\rm erg~s}^{-1},
\eeq
where $\dm$ is the accretion rate $\dMacc$ in units of $M_\odot$~s$^{-1}$. 
Ratio $L_0/\Lacc$ describes the efficiency of energy deposition
at the base of the jet; it is expected to be small. The only robust heating 
mechanism is neutrino-antineutrino annihilation, which gives
(Zalamea \& Beloborodov 2011)
\beq
   L_0\approx 10^{52}\, \dm^{9/4}\, \xms^{-4.8}
       \left(\frac{M}{3M_\odot}\right)^{-3/2} 
      {\rm erg~s}^{-1}.
\eeq
Here $\xms$ is the radius of the marginally stable Keplerian orbit in units
or $r_g$; it is determined by the spin parameter of the black hole $a$ and 
varies between $\xms=3$ for $a=0$ and $\xms=1/2$ for $a=1$. Optimistic 
assumptions regarding the black hole spin give $\xms\approx 1$ (which 
corresponds to $a=0.95$). In this case, heating peaks in a region of radius 
$r_0\sim 3 r_g\approx 10^6(M/3M_\odot)$~cm.
Note that the maximum $L_0\sim 10^{52}$~erg~s$^{-1}$ 
corresponds to a minimum $r_0\sim 10^6$~cm.

The thermal jet power $L_0$ is sensitive to the accretion rate and the black 
hole spin. It can vary by several orders of magnitude, which implies large
variations in observed luminosity (even when the beaming angle $\thb$ 
remains unchanged). The maximum $L_0\sim 10^{52}$~erg~s$^{-1}$ 
corresponds to the maximum achievable $\Ep$ of 10-15~MeV.


\section{Magnetically dominated jets}

It is possible that the energy output of the central engine is dominated 
by the Poynting flux, i.e. carried mainly by the magnetic field. Then the total 
energy per proton (cf. \Eq~\ref{eq:eta}) is given by
\beq
\label{eq:eta1}
    \eta=\etaB+\etath=const.
\eeq
Near the central engine the contribution of the thermal power is small, 
$\ep=\etath/\eta\ll 1$; it can increase at larger radii at the expense 
of the magnetic part $\etaB/\eta$.
The magnetic field is the main reservoir of energy and a large fraction of 
it must dissipate if the model aims to describe a bright, radiatively 
efficient burst.

Energy dissipated at radii $r<\RP$ must be thermalized into Planck radiation
with luminosity $\Lrad(\RP)=\etaP\dM c^2$ (where $\etaP=\etath[\RP]$) 
and with the average photon energy 
\beq
  \EavP=\etaP\,m_pc^2\,\left(\frac{n}{n_\gamma}\right)_{\RP}.
\eeq
At $r>\RP$ the photon-to-baryon ratio freezes out (as long as the main 
photon producer is the thermal plasma, i.e. nonthermal processes are less 
efficient). As dissipation continues in the Wien zone $r>\RP$,
$\Eav$ grows proportionally to $\etath$. 
The observed $\Ep$ is associated with $\EavW$ at $r\sim\RW$,
\beq
\label{eq:Ep_mag1}
   \EavW\approx \frac{\etaW}{\etaP}\,\EavP
                     =\frac{\etaW\, m_pc^2}{(n_\gamma/n)_{\RP}}, 
\eeq
where $\etaW=\etath(\RW)$. \Eq~(\ref{eq:Ep_mag1}) shows that $\Ep$ could 
be very high if the Planck zone is cold (which gives a small $n_\gamma/n$) 
and the Poynting flux dissipation is strong in the Wien zone (which gives a high 
$\etaW$). Such a jet would experience ``photon starvation'' --- the dissipated 
energy would be carried by a small number of photons with a high energy per 
photon, hence a high $\Ep$. This suggests that $\Ep$ in magnetically
dominated jets may exceed the maximum $\Ep$ of thermally dominated jets.
 
More specific estimates can be made as follows.
Magnetically dominated jets are expected to start with a modest Lorentz 
factor at the Alfv\'en radius and then gradually accelerate. In a simple 
self-similar model, the dissipated Poynting flux is distributed 
in comparable amounts between the thermal luminosity $\Lrad$ 
(dominated by radiation) and the proton kinetic energy flux, $L_{\rm kin}$. 
This means that $\etath\approx\Gamma$ and the thermal energy density in 
the fluid frame is approximately equal to the proton rest-mass density, 
\beq
\label{eq:U1}
    \Urad\approx 3kT\,n_\gamma\approx m_pc^2 n,
\eeq
which gives a simple relation between the jet temperature and the
photon-to-baryon ratio,
\beq
\label{eq:ratio_mag}
    \frac{n_\gamma}{n}\approx \frac{m_p}{3m_e\,\Th}.
\eeq
The freeze-out of $n_\gamma/n$ 
implies approximately constant temperature $\Th=kT/mc^2$ at $r>\RP$,
i.e. the jet temperature in the Wien zone remains approximately equal to $\TP$.
This fact may be understood in a slightly different way: conservation of 
the photon flux in the Wien zone implies 
$\Lrad\propto\Eav\approx 4\Gamma\,kT$; then the scaling 
$\Lrad\approx L_{\rm kin}\propto \Gamma$ requires $T=const=\TP$.

From \Eq~(\ref{eq:ThP}) we have
\beq
   \ThP\approx 0.01 \left(\frac{\RP}{10^{10}\rm~cm}\right)^{-1/6}
                                \left(\frac{\etaP}{30}\right)^{1/6}.
\eeq
Substitution of $\Th=\ThP$ into \Eq~(\ref{eq:ratio_mag}) gives the 
photon-to-baryon ratio for magnetically dominated jets,
\beq
    \frac{n_\gamma}{n}\approx 6\times 10^4.
\eeq
The observed $\Ep$ is close to $\EavW=4\GamW k\TW$, where 
$\GamW=\Gamma(\RW)$ and $\TW=T(\RW)\approx\TP$ is unchanged 
from the temperature at Planck radius $\TP$. This gives
\beq
\label{eq:Ep_mag}
     \Ep\approx 3\,\left(\frac{\GamW}{100}\right) {\rm~MeV}.
\eeq
In a radiatively efficient GRB dominated by photospheric emission, magnetic 
dissipation is expected to be nearly complete at subphotospheric radii (but see
\Sect~5.2). Then $\GamW$ may be comparable to the asymptotic value, 
$\GamW\simlt\eta$.

A monotonic dependence of $\Ep$ on  $\Gamma$ or $\eta$ is generally 
expected for magnetically dominated jets. Giannios (2012) estimated $\Ep$ 
assuming blackbody radiation $T=\Teff$ in the Wien zone $y\gg 1$.
He found $\Ep\propto\Gamma^{4/3}\eta^{1/3}$, which gives $\eta^{5/3}$ if 
$\Gamma\sim\eta$. Comparison of \Eq~(8) in Giannios (2012) with our 
\Eq~(\ref{eq:Ep_mag}) shows how the photon deficit in the Wien zone affects 
the observed spectral peak. $\Ep$ is significantly higher, in particular for 
moderate $\eta\sim 100$, and the scaling with $\eta$ is linear if $\GamW\sim\eta$.

$\Ep$ could be increased above the estimate~(\ref{eq:Ep_mag}) if magnetic 
dissipation is delayed so that heating is strongly suppressed at $r<\RP$.
Then $\Urad\ll n\,m_pc^2$ at the Planck radius, and the jet can be in the 
regime of a strong photon starvation. This regime would, however, require
a very cold central engine, to avoid
thermal photons transported by the jet from the center. 

In fact, the model described by \Eqs~(\ref{eq:U1})-(\ref{eq:Ep_mag}) 
already implies a relatively cold central object. It assumes that the density 
of heat advected from the center to $\RP$ is smaller than the heat generated 
by magnetic dissipation, and smaller than $n\,m_pc^2$.  
The model assumes that the initial thermal energy per baryon 
$\eta_{\rm th,0}$ becomes dynamically unimportant
(unable to accelerate the jet) before the temperature drops to $\TP$.
Only in this case the jet dynamics and photon number at $\RP$ are controlled 
by magnetic dissipation rather than by heat advected from the central engine.  
This condition requires an initial 
temperature 
\beq
   T_0\simlt \eta_{\rm th,0}\TP\approx 5\,\eta_{\rm th,0} {\rm ~keV}.
\eeq
Magnetically dominated models assume $\eta_{\rm th,0}\ll \eta$,
which leads to a strong upper bound on $T_0$. It would be inconsistent, 
for instance, with the collapsar model that invokes a hot accretion disk
with strong neutrino emission. A relatively cold central engine is expected 
in the proto-magnetar model (e.g. Metzger et al. 2011).


\section{Discussion}

\subsection{Regulation of $\Ep$}

Three characteristic radii are important for the formation of a photospheric GRB 
spectrum:

(1) Planck radius $\RP$ below which radiation is forced to have a Planck
spectrum. GRB jets have a well-defined temperature $\ThP\approx 0.01$ 
and Thomson optical depth $\tauP\approx 10^5$ at the Planck radius.
The typical value of $\RP$ is $10^{10}$~cm (\Eq~\ref{eq:RP4}).

(2) Wien radius $\RW$ below which $y\gg 1$ and radiation maintains a 
Wien (or Bose-Einstein) spectrum. Compton scattering enforces thermal 
coupling between the electrons and photons at $r<\RW$; however, the
radiation density can be far below the blackbody density $aT^4$.
The Wien zone $\RP<r<\RW$ occupies an extended range of optical depths, 
$\tauP>\tau>\tauW$, where $\tauW\sim 10^2$.
The existing transfer simulations indicate that the observed $\Ep$ is inherited
from the Wien zone.

(3) Photospheric radius $\Rph$ where optical depth $\tau=1$. 
Radiation is released around $\Rph$.
Thermal decoupling of plasma and radiation, $T_e>T_\gamma$, occurs
in the zone $\RW<r<\Rph$, and Comptonization changes 
the released spectrum from Wien to Band shape.

It is instructive to consider the role of entropy for the regulation of $\Ep$.
Entropy of GRB jets is strongly dominated by radiation. Entropy of Planck 
radiation is proportional to photon number. 
In a non-dissipative jet (whose entropy is conserved) radiation keeps a Planck 
spectrum even outside $\RP$, as this is consistent with constant photon
number --- the production of additional photons is not needed to maintain 
the thermal spectrum.

In dissipative jets, maintaining a Planck spectrum requires a growing photon 
number, which is possible as long as there are sufficiently fast processes 
producing photons. Such processes are guaranteed at $r<\RP$. Outside 
$\RP$, the thermal plasma becomes unable to supply new photons;
then the photon number freezes out and does not keep up anymore with 
the Planck value. This leads to the photon deficit $n_\gamma<\nP$
and the Wien spectrum between $\RP$ and $\RW$.
Photon deficit could be avoided 
if a significant fraction of the dissipated energy is injected in the form of 
nonthermal particles; then additional photons could be generated by synchrotron 
emission (\Sect~2.5; Vurm et al. 2012). However, the efficiency of nonthermal 
particle injection in the (very opaque) Wien zone is uncertain; in addition, 
the synchrotron photon supply is reduced by Bose condensation (Vurm et al. 2012).

In general, photospheric emission with efficiency $\ep=\Lrad/L$ and 
photon-to-baryon ratio $n_\gamma/n$ satisfies the following relation,
\beq
\label{eq:Ep_gen}
   \frac{\Ep}{\Ein}\approx \frac{\ep}{\ep_0} \frac{(n_\gamma/n)_0}{(n_\gamma/n)},
\eeq
where index ``0'' refers to the radius $r_0$ of the central engine; $\ep_0$ is 
the initial thermal fraction of the jet and $\Ein$ is given in \Eq~(\ref{eq:E0}).
Note that the photon number never decreases bellow its central value
(dissipation can only increase it), i.e. 
$Q=(n_\gamma/n)(n_\gamma/n)_0^{-1}\geq 1$.
\Eq~(\ref{eq:Ep_gen}) is applicable to both thermally dominated 
($\ep_0\approx 1$) and magnetically dominated ($\ep_0\ll 1$) jets;
it is valid regardless of dissipation or collimation mechanisms.

Non-dissipative flows have $Q=1$ and preserve a Planck spectrum 
everywhere in the opaque zone $r\ll\Rph$.\footnote{Transfer effects near the 
       photosphere $\Rph$ modify the observed spectrum into
       a multi-Doppler-shifted blackbody (Beloborodov 2010; Pe'er \& Ryde 2011).}
The resulting $\Ep=(\ep/\ep_0)\Ein$ is determined by the adiabatic cooling factor 
$\ep/\ep_0<1$, which is related to the radiative efficiency of photospheric 
emission $\ep$.

Dissipation affects $\Ep$ in two ways. Since heat is quickly passed to radiation, 
strong subphotospheric dissipation gives
a high radiative efficiency $\ep\sim 1$, offsetting the adiabatic cooling effect
(and in magnetic jets that start with a small thermal fraction 
$\ep_0\ll 1$, $\ep>\ep_0$ is possible). On the other hand, 
dissipation tends to generate photons, in particular in the Planck zone, where 
the photon number grows proportionally to the generated entropy. The observed 
$\Ep$ is sensitive to the photon production factor $Q$.
Typical observed GRBs are consistent with $Q\sim 10$ (Figure~1), suggesting 
significant dissipation in the Planck zone.
One can also see that bursts with record-high $\Ep\sim 10-20$~MeV
(Axelsson et al. 2012) must have $Q\sim 1$ (the minimum possible value) 
or be magnetically dominated near the central engine, $\ep_0\ll 1$. 

The lower bound on $\Ep$ is derived if one assumes unlimited
photon production that maintains detailed equilibrium up to the
photosphere. This would give a blackbody 
photospheric emission with $\Ep\approx\Emin$ given by \Eq~(\ref{eq:Emin}).
The condition $\Ep>\Emin$ gives a robust upper bound on the jet Lorentz 
factor,
\beq
   \Gamma<270\left(\frac{\Ep}{300 \rm~keV}\right)^{1/2} L_{\gamma,52}^{1/8} 
                          \,f_\pm^{1/4}\ep^{-1/4},
\eeq
where $f_\pm=1+n_\pm/n$ is the pair loading factor. 

In the thermally dominated limit, $\etath\gg\eta_B$, the maximum value for 
$\Ep$ is set by the initial temperature near the central engine 
(\Eq~\ref{eq:Epmax}).
A plausible scenario invokes comparable contributions of the initial thermal
energy and magnetic field to the jet power, $\etath\sim\etaB$. 
Then the maximum $\Ep\sim 10$~MeV expected for the pure thermal jet
can be increased by a factor $\sim 2$ by dissipation of additional 
magnetic energy that has been transported by the Poynting flux to 
$r>\RP$. This picture is consistent with the observed distribution of $\Ep$,
which cuts off at about 20~MeV.

In the magnetically dominated limit, $\etaB\gg\etath$, the initial temperature 
plays no role. In this case, the thermal output of the central engine is 
negligible and heat/photons are gradually generated in the expanding jet as 
a result of magnetic dissipation. The simple self-similar model gives 
a unique $n_\gamma/n\sim 6\times 10^4$ at the Planck radius, and the 
observed $\Ep$ is proportional to the Lorentz factor of the jet 
(\Eq~\ref{eq:Ep_mag}).

\subsection{Magnetic dissipation}

The rate of magnetic dissipation is hard to calculate from first principles;
e.g., the reconnection rate depends on the field structure in the jet.
Theory is easily reconciled with observations if most of 
dissipation occurs at subphotospheric radii, $\Rdiss<\Rph$.
The MeV peak is observed to carry most of the GRB energy, and we argued 
in \Sect~1 that it emerges from the photosphere, which implies that most 
of electron heating occurs at $r<\Rph$. 

The photospheric radius steeply decreases with $\Gamma$, 
$\Rph\propto \Gamma^{-3}$, while $\Rdiss$ increases with $\Gamma$,
e.g. in the magnetic dissipation model of Drenkhahn \& Spruit (2002).
Then the condition $\Rdiss<\Rph$ typically means $\Rdiss\ll\Rph$ 
(as $\Rdiss\sim\Rph$ would require fine-tuning of $\Gamma$).
This feature of magnetic dissipation is avoided only in models with 
reconnection rate suppressed at $\tau\gg 1$ and quickly increasing at 
$\tau\sim 1$ (Mckinney \& Uzdensky 2012).

If dissipation peaks at $r<\RW$, it must influence $\Ep$. In this scenario, 
the dissipation of the energetically dominant Poynting flux is hidden 
at high optical depths, and it does not dominate the heating at $r\simlt\Rph$ 
where the Wien spectrum is Comptonized into a Band shape.

The remaining source of energy in the Comptonization zone is internal bulk 
motions (which may be initiated by magnetic dissipation at smaller radii).
In particular, collisional dissipation, which provides significant electron heating,
is expected to peak at moderately subphotospheric radii $r\simlt\Rph$. 
It begins at $R_n\sim (\sigma_n/\sT)\Rph$, where $\sigma_n\approx\sT/20$ is 
the nuclear cross section, and converts a large fraction of the jet energy to 
electron heat and nonthermal $e^\pm$ pairs.

Models that invoke strong magnetic dissipation extending through the photosphere 
may be consistent with observations if the released energy is given to protons, 
and electrons receive 
energy from protons via Coulomb collisions. Coulomb coupling is efficient only 
below the photosphere; therefore, electrons receive and radiate the dissipated 
energy at subphotospheric radii $r<\Rph$.
(At radii $r>\Rph$ most of the proton heat is not radiated --- it is lost to 
adiabatic cooling and converted to the bulk kinetic energy of the jet.) 
This scenario is a ``magnetically powered'' variation of the collisional 
mechanism. The heating of protons results in bright
emission due to $e$-$p$ Coulomb energy exchange. In addition, it may lead to 
inelastic nuclear $p$-$p$ collisions; this process will inject $e^\pm$ pairs 
with Lorentz factors $\gamma\sim m_\pi/m_e\sim 300$ and generate an 
extended high-energy tail of the GRB spectrum. This model is similar to that 
of  Beloborodov (2010), which examined $n$-$p$ collisions.  The only 
difference is that here the source of heat is the magnetic field 
instead of the relative streaming of the neutron and proton components of the jet.

\subsection{Variations in $\Ep$}

Various reported correlations between $\Ep$, $\Lrad$, and $\Gamma$ may
be compared with theoretical predictions for photospheric emission (e.g.
Giannios 2012; Fan et al. 2012). We argued in this paper that one should not 
make blackbody assumptions in this analysis; instead, one should examine 
the entire expansion history of the jet.
Besides the energy output of the central engine, $L_0$, and its thermal fraction 
$\ep_0$, the observed emission is controlled by two factors: beaming $L/L_0$ 
and the photon production factor $Q$ (Figure~1). 
Photon production depends on dissipation in the Planck zone
due to collimation shocks, or possibly due to magnetic reconnection.

A positive correlation between $\Ep$ and the burst 
luminosity is expected for dissipative jets; e.g. 
thermally dominated jets with fixed $r_0=const$ and 
beaming angle $\thb=const$ would have $\Ep$ scaling as $\Lrad^{1/4}$.
The $\Ep$-$\Lrad$ correlation steepens in the presence of magnetic 
dissipation if brighter bursts have higher $\Gamma$.
The relation between $L$ and $\Ep$ can also be affected by a correlation 
between the jet opening angle $\theta_b$ and the photon production factor $Q$.

Low $\Ep$ is naturally associated with a large $Q$. 
It is also associated with a low true luminosity $L_0$ (and the 
correspondingly low central temperature $T_0$). 
A strong collimation can boost the apparent luminosity $L$ from a low 
$L_0$, however it cannot increase $\Ep$ (\Sect~3). 

The dependence of $\Ep$ on luminosity $L$ and photon number
helps understand the pattern of $\Ep$ variations in individual 
pulses of GRB light curves. Although a tracking behavior $\Ep(L)$ is 
expected, the presence of the second parameter can lead 
to significant deviations. In particular, the observed high $\Ep$ at the 
beginning of a pulse (before its luminosity reaches maximum) may be
explained as emission with a low photon-to-baryon ratio. 
As the pulse progresses, $n_\gamma/n$ grows until
the normal tracking behavior is established. In addition, the observed 
$L$ depends on the beaming factor, which may vary during the burst; 
this may also contribute to the deviations from the tracking behavior 
$\Ep(L)$.

\vspace{0.2in}
I thank Amir Levinson and Indrek Vurm for useful comments on the manuscript.
This work was supported by NSF grant AST-1008334.


\begin{appendix}

\section{Photon production rates}

The rate of photon production by the double Compton effect has been 
extensively discussed in the literature (e.g. Thorne 1981; Lightman 1981;
Pozdnyakov et al. 1983; Svensson 1984; Chluba et at. 2007).
Scattering of mono-energetic photons of energy $h\nu_0$ and density 
$n_\gamma$ on cold electrons of density $n$ produces secondary photons 
$h\nu$ with the following differential rate,
\beq
\label{eq:rateDC1}
   \frac{d\dnDC}{d\ln\nu}=\frac{4\alpha}{3\pi}\,n\,n_\gamma\,\sT\,c\,
                               \left(\frac{h\nu_0}{m_ec^2}\right)^2,
\eeq
where $\alpha=e^2/\hbar c=1/137$ is the fine structure constant.
\Eq~(\ref{eq:rateDC1}) is valid when the secondary photon has energy 
$h\nu\ll h\nu_0$. Extension to $\nu\sim\nu_0$ was discussed by Gould (1984); 
it gives only a small correction to the total photon production rate $\dnDC$, 
as $\dnDC$ has a flat distribution over $\ln\nu$ and most 
photons are emitted with $\nu\ll\nu_0$.

For primary photons $h\nu_0$ with a given spectrum,
rate (\ref{eq:rateDC1}) should be averaged over the spectrum. For a Bose-Einstein
radiation with temperature $T$, this gives
\beq
\label{eq:rateDC2}
     \frac{d\dnDC}{d\ln x}=\frac{4\alpha}{3\pi}\,n\,n_\gamma\,\sT\,c\,\overline{x_0^2}\,
       \Th^2, \qquad x_0=\frac{h\nu_0}{kT},  \qquad x=\frac{h\nu}{kT}, 
            \qquad \Th=\frac{kT}{m_ec^2},
\eeq
where $\overline{x_0^2}=12\zeta(5)/\zeta(3)\approx 10.35$ for a Planck spectrum and 
$\overline{x_0^2}=12$ 
for a Wien spectrum; photons with $x_0\simgt 1$ make the dominant contribution.

The net rate $\dnDC$ is obtained by integrating \Eq~(\ref{eq:rateDC2}) over 
$x$ from a minimum value $\xmin$ to $x\sim 1$. Here $\xmin$ is a
minimum energy of produced photons that avoid absorption and get a chance 
to be Comptonized to the Wien peak (Thorne 1981; Lightman 1981). 
It is determined by equating the absorption rate $\tabs^{-1}=\aDC\, c$ 
(where $\aDC$ is the absorption coefficient due to inverse double Compton 
effect) and the Comptonization rate $\tIC^{-1}$ (where $\tIC$ is the time 
it takes to Comptonize the photon energy by a factor of 2).
Absorption coefficient $\aDC$ is found from Kirchhoff's law for
Rayleigh-Jeans radiation $\aDC 8\pi\nu^2 kT/c^2=h\nu\,d\dnDC/d\nu$.
Comptonization rate at non-relativistic temperatures is 
$\tIC^{-1}\approx 4\Th n\sT c$. This gives,
\beq
   \xmin=\left(\frac{\pi}{3}\,\alpha\,\lbar^3n_\gamma\,\overline{x_0^2}\,\Th^{-2}\right)^{1/2}
            =\left(\frac{8}{\pi}\,\alpha\,\zeta(5)\,\Th \right)^{1/2}\approx 0.14\Th^{1/2},
\eeq
where $\lbar=\hbar/m_ec$ is Compton wavelength. We have used the relation 
$\lbar^3 n_\gamma \overline{x_0^2}=(24/\pi^2)\zeta(5)\,\Th^3$ for blackbody 
radiation, where $\zeta(5)\approx 1.037$. Integration of \Eq~(\ref{eq:rateDC2}) gives
\beq
    \dnDC=\chi\,n\,n_\gamma\,\sT\,c\,\Th^2,   \qquad
                \chi=\frac{4\alpha}{3\pi}\,\overline{x_0^2}\,\ln\xmin^{-1}.
\eeq
This equation assumes that the scattering electrons are cold in the sense that 
$\Th\ll 1$. In the case of a hot plasma it should be multiplied by a correction 
factor that has been obtained by Svensson (1984) (see also Chluba et al. 2007). 
The correction factor is given by
\beq
   g_{\rm DC}=\left(1+13.91\Th+11.05\Th^2+19.9\Th^3\right)^{-1}.
\eeq
At the boundary of the Planck zone we find $\ThP\approx 10^{-2}$ (\Sect~2.2);
then $\chi\approx 0.1$.

Photon production by bremsstrahlung is given by
(e.g. Illarionov \& Sunyaev 1975; Thorne 1981; Pozdnyakov et al. 1983),
\beq
\label{eq:rateB}
     \frac{d\dnB}{d\ln x}=\left(\frac{2}{\pi}\right)^{3/2}\alpha\,n^2\,\sT\,c\,
       \Th^{-1/2}\,\ln\frac{2.2}{x}.
\eeq
Relativistic corrections (e.g. Svensson 1984) are small 
for temperatures of interest here. 
Comparison with \Eq~(\ref{eq:rateDC2}) shows that at relevant  
temperature $\ThP\simgt 0.01$ and photon-to-baryon ratio 
$n_\gamma/n\sim 10^5$, the bremsstrahlung emissivity is smaller 
or comparable to the double Compton emissivity. 
Integration of \Eq~(\ref{eq:rateB}) over $\ln x$ from $\ln\xmin$ to 
$\ln x\sim 0$ gives
\beq
    \dnB=\xi\,n^2\,\sT\,c\,\Th^{-1/2},   \qquad 
         \xi\approx \left(\frac{2}{\pi}\right)^{3/2} \alpha \left(\ln\xmin^{-1}\right)^2.
\eeq
Near the boundary of the Planck zone $\xi\approx 0.06$.

\end{appendix}


\end{document}